\documentclass[twocolumn]{aastex63}

\received{\today}
\revised{\today}
\accepted{\today}
\submitjournal{ApJ}

\shorttitle{uGMRT observations of AT\,2018cow}
\shortauthors{Nayana et al.}
\graphicspath{{./}{figures/}}

\begin{document}

\title{uGMRT observations of a Fast and Blue Optical Transient - AT\,2018cow}

\correspondingauthor{Nayana A.J.}
\email{nayan89deva@gmail.com}

\author[0000-0002-0844-6563]{Nayana A.J.}
\affiliation{Department of physics, United Arab Emirates University, Al-Ain, UAE, 15551.}
\affiliation{National Centre for Radio Astrophysics, Tata Institute of Fundamental Research, PO Box 3, Pune, 411007, India.}
\author[0000-0002-0844-6563]{Poonam Chandra}
\affiliation{National Centre for Radio Astrophysics, Tata Institute of Fundamental Research, PO Box 3, Pune, 411007, India.}



\begin{abstract}

We present low-frequency radio observations of a fast-rising blue optical transient (FBOT), AT\,2018cow, with the upgraded Giant Metrewave Radio Telescope (uGMRT). Our observations span $t =$ 11 $-$ 570 days post-explosion and a frequency range of 250 $-$ 1450 MHz. The uGMRT light curves are best modeled as synchrotron emission from an inhomogeneous radio-emitting region expanding into an ionized medium. However, due to the lack of information on the source covering factor, which is a measure of the degree of inhomogeneity, 
we derive various parameters assuming the source covering factor to be unity. These parameters, hence, indicate limits on the actual values in an inhomogeneous model. We derive the lower limit of the shock radius to be $R \sim$ (6.1$-$14.4) $\times$ 10$^{16}$ cm at $t =$ 138$-$257 days post-explosion. We find that the fast-moving ejecta from the explosion are moving with velocity $v$ $>$ 0.2c up to $t =$ 257 days post-explosion. The upper limits of the mass-loss rate of the progenitor are $\dot{M}$ $\sim$ (4.1$-$1.7) $\times$ 10$^{-6}$ $M_{\odot}$\,yr$^{-1}$ at (19.3$-$45.7) years before the explosion for a wind velocity $v_{\rm w}$ = 1000 km\,s$^{-1}$. These $\dot{M}$ values are $\sim$ 100 times smaller than the previously reported mass-loss rate at 2.2 years before the explosion, indicating an enhanced phase of the mass-loss event close to the end-of-life of the progenitor. Our results are in line with the speculation of the presence of a dense circumstellar shell in the vicinity of AT\,2018cow from previous radio, ultra-violet, and optical observations.    

\end{abstract}

\keywords{FBOT: general --- FBOT: AT\,2018cow --- radiation mechanisms: non-thermal --- circumstellar matter --- radio continuum: general}


\section{Introduction} \label{sec:intro}
Fast Blue Optical transients (FBOTs) are a class of transients characterised by high optical luminosity ($\geq$10$^{43}$ erg\,s$^{-1}$), fast rise time ($t <$12 days) and blue colours \citep{drout2014,rest2018}. The observational properties of all FBOTs are not in-line with traditional supernova models \citep{drout2014,pursiainen2018,rest2018}. The high peak luminosity and rapid rise time of FBOTs are difficult to explain with the radioactive decay of $^{56}$Ni unless a very large Ni mass is assumed \citep{drout2014}. There exist two categories of models in the literature to explain the energy of FBOTs, which is not due to radioactive decay. One is the interaction of the explosion shock with the surrounding medium \citep{chevalier2011,balberg2011,ginzburg2014} and the other is the power supplied by a central compact object \citep{yu2013,metzger2014,hotokezaka2017}.
A detailed study of the properties of FBOTs as a group is limited 
because of the small number of known FBOTs, which is due to the low
discovery rate.

AT\,2018cow was discovered on 2018 June 16.44 UT \citep{smartt2018} with the Asteroid Terrestrial-Impact Last Alert System \citep[ATLAS;][]{tonry2018} in the dwarf star-forming galaxy CGCG\,137$-$068 at a distance of 66 Mpc \citep{prentice2018}. The source was not detected by All Sky Automatic Survey for Supernovae (ASAS-SN) on 2018 Jun 15.14 UT \citep{prentice2018}, tightly constraining the time of event. We assume 2018 Jun 16.44 UT to be the time of the event {\bf($t =0$)} and all times ($t$) are mentioned with respect to this time throughout the paper.

AT\,2018cow has been classified as an FBOT \citep{perley2019} and is the first FBOT detected in the local Universe. Other than AT\,2018cow there are only two FBOTs for which detailed multi-wavelength follow-up observations exist; CSS161010 \citep{deanne2020} and ZTF18abvkwla \citep{ho2020}. 
AT\,2018cow displayed several peculiar characteristics in its early evolutionary phase; rapid rise to peak \citep[$t_{\rm peak} \leq$ 3 days;][]{fremling2018,prentice2018}, high peak optical luminosity ($L_{\rm peak} \sim$ 10$^{44}$ erg\,s$^{-1}$) followed by a relatively fast decay \citep[$t^{-2.5}$;][]{perley2019}, initially featureless optical spectrum followed by broad short-lived spectral features \citep{perley2019,prentice2018,margutti2019}, luminous and variable X-ray emission \citep{margutti2019} and bright sub-mm radio emission \citep{ho2019}. Radio emission was detected from AT\,2018cow at various frequencies from 1.4 GHz {\bf \citep{nayana2018}} to 34 GHz \citep{margutti2019,dobie2018} including Very long baseline Interferometry (VLBI) observations \citep{mohan2019,bietenholz2019}. 
There exist various progenitor scenarios in the literature attempting to explain the observed properties of AT\,2018cow. These include stellar collapse leading to a compact object like a black hole or magnetar \citep{margutti2019}, a tidal disruption event \citep[TDE;][]{perley2019}, a merger of two white-dwarfs \citep{lyutikov2019}, a jet driven by the collision of accreting neutron star and a star \citep{soker2019} and a failed explosion of a blue supergiant \citep{margutti2019} and an explosion of a luminous
blue variable to an inhomogeneous CSM \citep{rivera2018}.

Regardless of the actual nature of the explosion, the radio emission from AT\,2018cow is understood to be from the fastest ejecta interacting with the surrounding medium \citep{ho2019,margutti2019}. \cite{ho2019} derived a shock velocity of $v \sim$ 0.13 c in a medium of density $n_{\rm e}$ = 3 $\times$ 10$^{5}$ cm$^{-3}$ from early (up to day 81) sub-mm observations. The authors invoke a model in which the transient explodes in a dense CSM shell of radius $\sim$ 1.7 $\times$ 10$^{16}$ cm to explain multi-epoch radio observations. \cite{margutti2019} presented multi-frequency Karl G. Jansky Very Large Array (JVLA) observations of AT\,2018cow during $t= $ 82 to 150 days and model the radio emission as
coming from a shock of velocity $v=0.1$c interacting with a dense environment. While the early radio observations probe the density and shock properties in the immediate vicinity of the transient, low-frequency observations at late times trace the environmental properties at larger radii.

We present the lowest frequency radio observations of AT\,2018cow, carried out with the upgraded Giant Metrewave Radio telescope (uGMRT) covering a frequency range 
$\sim 250-1450$ MHz during 11$-$570 days post-explosion. We explain the uGMRT observations and data reduction in \S \ref{sec:obs}. The radio light curves and modeling are presented in \S \ref{sec:radio-lc-modeling}. The properties of the shock and mass-loss rate of the progenitor are inferred in \S \ref{sec:results and discussion}. We compare the properties of AT\,2018cow with other energetic transients in \S \ref{sec:comparison} and summarize our results in \S \ref{sec:summary}.

\section{UGMRT Observations and Data Reduction}
\label{sec:obs}
 We carried out uGMRT observations of AT\,2018cow from 2018 Jun 27.8 ($t$ = 11 days) till 2020 Jan 7.1 ($t$ = 570 days). The observations were carried out in  the uGMRT band 5 (1050$-$1450 MHz), band 4 (550$-$950 MHz)  and band 3 (250$-$500) MHz.
 The data were recorded in 2048 frequency channels covering a bandwidth of 400 MHz in the bands 5 and 4 and 200 MHz in the band 3 with an integration time of 10 seconds in the total intensity mode. 3C286, 3C147, and 3C48 were used as the flux density and bandpass calibrators. The data were analysed using the Common Astronomy Software Application package \citep[CASA;][]{mcmullin2007}. The data were flagged and calibrated using standard CASA tasks. The calibrated visibility data were imaged using CASA task TCLEAN. A few rounds of phase only self-calibration were performed to improve the image quality. The source flux density and errors are determined by fitting a two dimensional Gaussian to the emission using the GAUSSFIT\footnote{https://casa.nrao.edu/casadocs/casa-5.4.1/image-cube-visualization/regions-in-the-viewer} tab available in CASA region manager panel. The free parameters of the fit are position angle, centre, major and minor axis of the Gaussian. The dimensions of the fitted Gaussian are consistent with a point source at all three frequency bands. The details of uGMRT observations and flux densities at central frequencies 1.25, 0.75 and 0.40 GHz in bands 5, 4 and 3, respectively, are presented in Table \ref{tab:gmrt}. In addition to the gaussfit errors as given in Table \ref{tab:gmrt}, we add a 10\% systematic error in the band 5 and 4 and 15\% in the band 3 to account for the calibration uncertainties.

\begin{deluxetable}{cccc}
\tablenum{1}
\tablecaption{Details of the uGMRT observations of AT\,2018cow. \label{tab:gmrt}}
\tablewidth{0pt}
\tablehead{
\colhead{Date of observation} & \colhead{Age} & \colhead{Frequency} & \colhead{Flux density}  \\
\colhead{(UT)} & \colhead{(Day)} & \colhead{(GHz)} & \colhead{($\mu$Jy)}
}
\startdata
2018 Jun 27.83    & 11.39     & 1.25   &    $<$75           \\
2018 Jul 05.83    & 19.39     & 1.25   &    $<$69           \\
2018 Jul 16.47    & 30.03     & 1.25   &   110$\pm$20    \\
2018 Aug 12.71    & 57.27     & 1.25   &   350$\pm$82    \\
2018 Sep 07.58    & 83.14     & 1.25   &   810$\pm$65    \\
2018 Sep 24.77    & 100.33    & 1.25   &   872$\pm$58    \\
2018 Nov 02.26    & 138.82    & 1.25   &   2882$\pm$69    \\
2018 Dec 06.77    & 173.33    & 1.25   &   992$\pm$34    \\
2019 Feb 04.77    & 233.33    & 1.25   &   882$\pm$26    \\
2019 Apr 21.77    & 309.33    & 1.25   &   438$\pm$23    \\
2019 Sep 14.77    & 455.33    & 1.25   &   169$\pm$20    \\
2020 Jan 06.33    & 568.89    & 1.25   &   93$\pm$30    \\
2018 Sep 30.34    & 105.90    & 0.75   &   453$\pm$21    \\
2018 Nov 02.42    & 138.98    & 0.75   &   572$\pm$19    \\
2018 Dec 08.77    & 175.33    & 0.75   &   755$\pm$27    \\
2019 Jan 29.77    & 227.33    & 0.75   &   626$\pm$33    \\
2019 Apr 22.77    & 310.33    & 0.75   &   458$\pm$35    \\
2019 Sep 14.77    & 455.33    & 0.75   &   303$\pm$39    \\
2020 Jan 07.23    & 569.79    & 0.75   &   131$\pm$29    \\
2018 Nov 04.39    & 140.95    & 0.40   &   342$\pm$70    \\
2018 Dec 07.25    & 173.81    & 0.40   &   518$\pm$50    \\
2019 Feb 05.77    & 234.33    & 0.40   &   564$\pm$65    \\
2019 Apr 22.77    & 310.33    & 0.40   &   532$\pm$65    \\
2019 Sep 14.77    & 455.33    & 0.40   &   410$\pm$40    \\
2020 Jan 07.06    & 569.62    & 0.40   &   304$\pm$59    \\
\enddata
\tablecomments{The age is calculated assuming 2018 June 16.44 (UT) as the time of event (see \S \ref{sec:intro}). The listed uncertainities are only statistical uncertainities. There is also a 10 - 15 \% systematic uncertainity (10\% for 1.25 and 0.75 GHz flux densities and 15\% for 0.40 GHz flux densities) above that.}
\end{deluxetable}

\begin{figure*}
\begin{centering}
\includegraphics[scale=0.44]{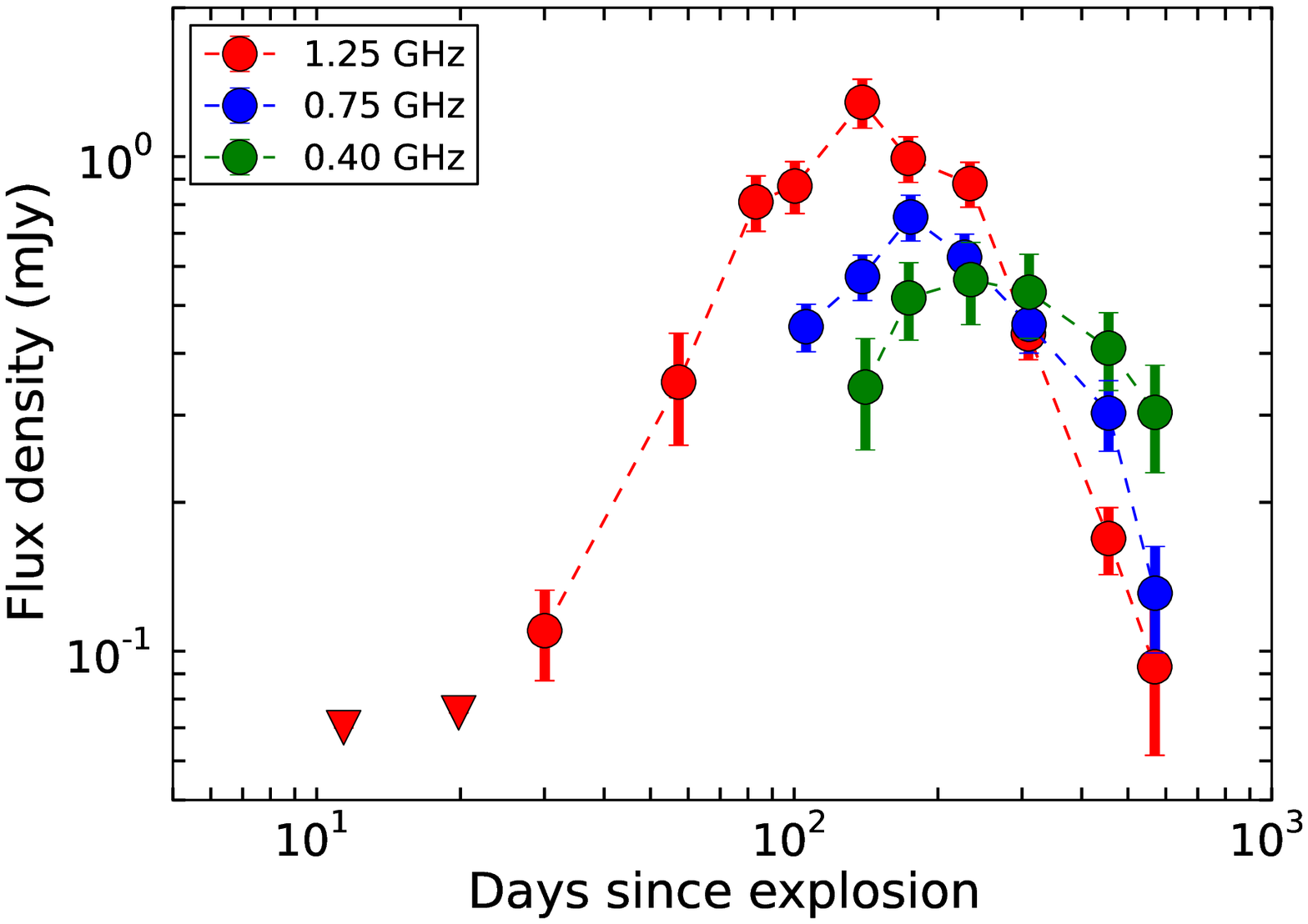} 
\includegraphics[scale=0.44]{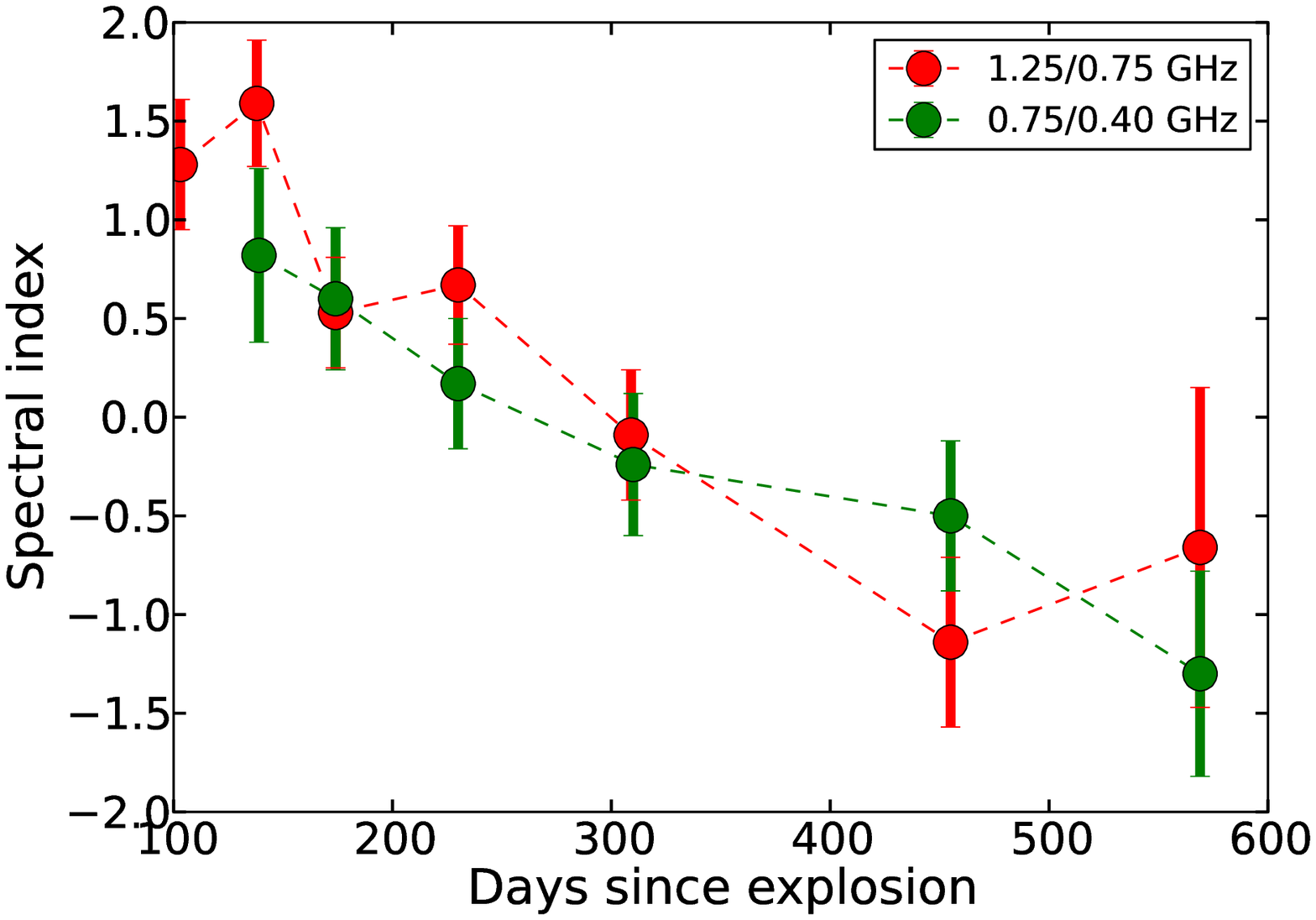}
\caption{Left panel: The uGMRT light curves of AT\,2018cow at 1.25, 0.75 and 0.40 GHz. Right panel: The near simultaneous spectral indices of AT\,2018cow between frequencies 1.25/0.75 GHz and 0.75/0.40 GHz at multiple epochs.}
\end{centering}
\label{fig:lc-gmrt}
\end{figure*}

\section{Radio light curves and modelling}
\label{sec:radio-lc-modeling}
We present the uGMRT light curves at 1.25, 0.75 and 0.40 GHz in Fig \ref{fig:lc-gmrt} (left panel). The transition from optically thick to thin regime is seen in the light curves at all frequencies. The peak spectral luminosity at 1.25 GHz is $L_{\nu \, \rm peak} =$ 5.3 $\times$ 10$^{27}$ erg\,s$^{-1}$\,Hz$^{-1}$. The spectral indices between 1.25/0.75 GHz and 0.75/0.40 GHz are shown in Fig \ref{fig:lc-gmrt} (right panel). The spectral index $\alpha$ ($F_{\nu} \propto \nu^{\alpha}$) in the optically thin regime approaches $\alpha =$  $-$0.80 $\pm$ 0.50 at $t >$ 400. The power-law index of the relativistic electron distribution $p$ ($N(E) \propto E^{-p}$) is related to $\alpha$ as $\alpha$ = $-(p-1)/2$, giving $p$ = 2.60 $\pm$ 1.00.

We model radio emission from AT\,2018cow as synchrotron radiation from a shock created due to the interaction of the ejecta with the surrounding medium. Initially the radio emission could be suppressed by free-free absorption (FFA) by the ionized external medium \citep{chevalier1982} or synchrotron self-absorption \citep[SSA;][]{chevalier1998}. 
The radio flux density, $F(\nu,t)$ in the FFA model is \citep{chevalier1982,weiler2002} 
\begin{equation}
F(\nu,t)=K_{1} \left( \frac{\nu}{5\hspace{0.1 cm} \rm GHz}\right)^{\alpha} \left( \frac{t}{10\, \rm day}\right)^{\beta}e^{-\tau_{\rm FFA}(\nu,t)}
\end{equation}
\begin{equation}
\tau_{\rm FFA}(\nu,t)=K_{2}\left(\frac{\nu}{5\hspace{0.1 cm}\rm GHz}\right)^{-2.1}\left(\frac{t}{10\, \rm day}\right)^{\delta}
\end{equation}
Here $\tau_{\rm FFA}$ denotes the free-free optical depth and $K_{1}$, $K_{2}$ are the flux density and optical depth normalizations respectively. $\alpha$, $\beta$ and $\delta$ denote the spectral and temporal indices of the radio flux densities and temporal index of optical depth, respectively. The radio flux density in the SSA model is \citep{chevalier1998,weiler2002}
\begin{equation}
F(\nu,t)=K_{1} \left( \frac{\nu}{5\, \rm GHz}\right)^{2.5} \left( \frac{t}{10\, \rm day}\right)^{a} \left(1- e^{-\tau_{\rm SSA}(\nu,t)} \right) 
\end{equation}
\begin{equation}
\tau_{\rm SSA}(\nu,t)=K_{2}\left(\frac{\nu}{\rm 5\, GHz}\right)^{-(p+4)/2}\left(\frac{t}{10\,\rm day}\right)^{-(a+b)}
\end{equation}
Here $\tau_{\rm SSA}$ denotes the SSA optical depth. $K_{1}$, $K_{2}$ are the flux density and optical depth normalizations, respectively. $a$, $b$ denote the temporal indices of the flux density in the optically thick and thin phases, respectively. 

\begin{figure*}
\includegraphics[scale=0.53]{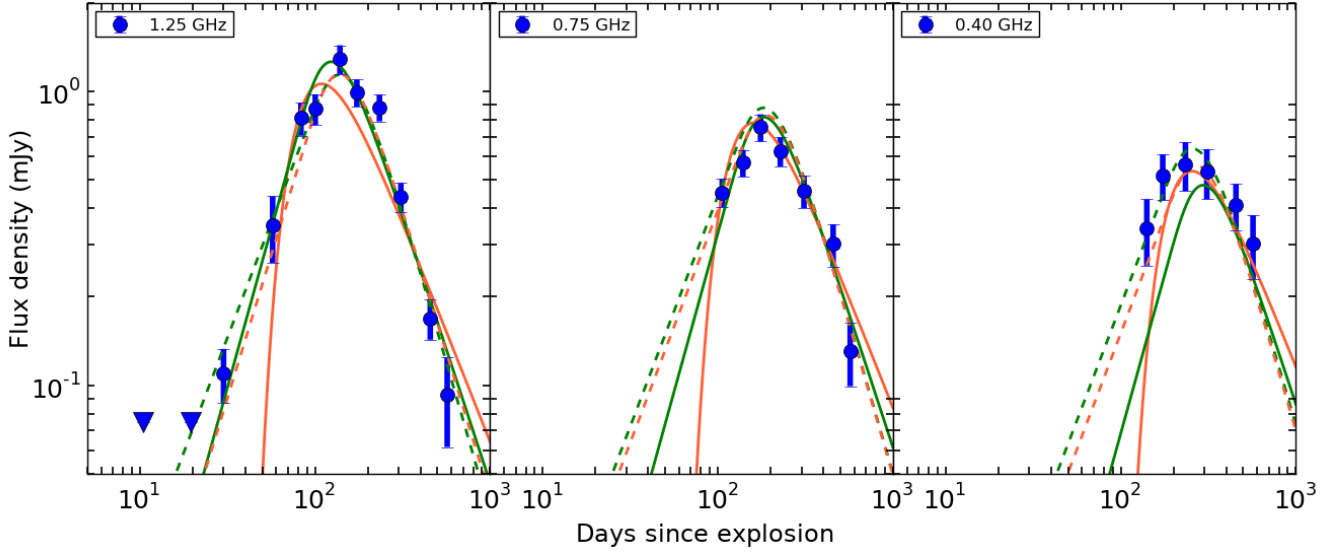}
\caption{The uGMRT light curves of AT\,2018cow at 0.40, 0.75 and 1.25 GHz frequencies. The green and red solid lines denote the best fit SSA and FFA models respectively. The green and red dotted lines denote the best fit inhomogeneous SSA and FFA models respectively.}
\label{fig:lc-fit-subplot}
\end{figure*}

We perform a combined fit where the data at all $\nu$ and $t$ are fit simultaneously with both FFA and SSA models using a chi-square minimization algorithm, curve$_{-}$fit in python-scipy \footnote{https://docs.scipy.org/doc/scipy/reference/}. $K_{1}$, $K_{2}$, $\alpha$, $\beta$ and $\delta$ are the free parameters in the FFA model and $K_{1}$, $K_{2}$, $a$, $b$ and $p$ are the free parameters in the SSA model. The best fit models along with the observed data are shown in Fig \ref{fig:lc-fit-subplot}. The best fit parameters in the FFA model are $K_{1}=32.11\pm19.34$, $K_{2}=24.14\pm22.98$, $\alpha = -0.53\pm0.30$, $\beta=-1.51\pm0.21$ and $\delta=-2.81\pm0.44$. The best fit SSA parameters are $K_{1}=0.26\pm0.14$, $K_{2}=366.05\pm286.71$, $a=2.16\pm0.23$, $b=1.82\pm0.19$ and $p=2.10\pm0.48$. The SSA model gives a better fit to the data ($\chi_{\nu}^{2}$ = 2.6) than the FFA model ($\chi_{\nu}^{2}$ = 4.7).  
The average optically thick spectral index is $\alpha$ = 1.23 $\pm$ 0.36, flatter than the standard SSA or FFA values. This can be attributed to the inhomogeneities in the emitting region or CSM \citep{rybicki1986,chandra2019,bjornsson2017,weiler2002}.  
In the FFA model, a clumpy CSM could lead to inhomogeneous absorption and the radio flux density in this case is \citep{weiler2002}
\begin{equation}
\label{eqn:ffa-standard}
F(\nu,t)=K_{1}^{'} \left( \frac{\nu}{5\, \rm GHz}\right)^{\alpha^{'}} \left( \frac{t}{10\, \rm day}\right)^{\beta^{'}} \left( \frac{1- e^{-\tau_{\rm FFA}^{'}(\nu,t)}}{\tau_{\rm FFA}^{'}} \right) 
\end{equation}
\begin{equation}
\label{eqn:ffa-tau}
\tau^{'}_{\rm FFA}(\nu,t)=K_{2}^{'}\left(\frac{\nu}{\rm 5\, GHz}\right)^{-2.1}\left(\frac{t}{10\,\rm day}\right)^{\delta^{'}}
\end{equation}
$K_{1}^{'}$ and $K_{2}^{'}$ denote the flux density and optical depth normalization respectively. The term (1$-$e$^{-\tau_{\rm FFA}^{'}}$)/$\tau_{\rm FFA}^{'}$ describes the absorption due to a clumpy CSM where $\tau_{\rm FFA}^{'}$ is the FFA optical depth. $\alpha^{'}$, $\beta^{'}$ and $\delta^{'}$ denotes the spectral and temporal indices of the radio flux densities and temporal index of optical depth respectively.  

An inhomogeneous synchrotron emitting region can be created due to the inhomogeneous distribution of relativistic electrons and/or magnetic fields. The inhomogeneity can be charaterized by a source covering factor, $f_{\rm B,cov}$ that describes the variation of the average magnetic field strength over the projected source surface \citep{bjornsson2017}. The covering factor gives rise to a range of optical depths over the source and hence broadens the spectrum. $P(B) \propto B^{\rm -a}$, is the probability to find a magnetic field of strength $B$. $f_{\rm B, cov}$ can be parametrized as $f_{\rm B,cov}$ $\approx$ $f_{\rm B_{0},cov} (B/B_{0})^{1-a}$, for $B_{\rm 0} < B < B_{\rm 1}$, where $f_{\rm B_{0}, cov}$ is the source covering factor for a magnetic field $B_{0}$. The observed spectrum will have three regions defined by the synchrotron self-absorption frequency ($\nu_{\rm abs}$); a standard optically thick region where $\nu < \nu_{\rm abs}(B_{\rm 0})$, a standard optically thin region $\nu > \nu_{\rm abs}(B_{\rm 1})$, and a transition region where $\nu_{\rm abs}(B_{\rm 0}) < \nu < \nu_{\rm abs}(B_{\rm 1})$. The radio flux density in the inhomogeneous SSA model is \citep{chandra2019,bjornsson2017}
\begin{equation}
\label{eqn:ssa-standard}
F(\nu,t)=K_{1}^{'} \left( \frac{\nu}{5\, \rm GHz}\right)^{\alpha^{''}} \left( \frac{t}{10\, \rm day}\right)^{a^{'}} \left(1- e^{-\tau_{\rm SSA}^{'}(\nu,t)} \right) 
\end{equation}
\begin{equation}
\label{eqn:ssa-tau}
\tau_{\rm SSA}^{'}(\nu,t)=K_{2}^{'}\left(\frac{\nu}{\rm 5\, GHz}\right)^{-(\alpha^{''} + \frac{p^{'}-1}{2})}\left(\frac{t}{10\,\rm day}\right)^{-(a^{'}+b^{'})}
\end{equation}

Here $\alpha^{''}$ is the spectral index in the transition region. $K_{1}^{'}$ and $K_{2}^{'}$ denote the flux density and optical depth normalization respectively.
In an inhomogeneous model, it is assumed that the locally emitted spectrum is that of the standard synchrotron. However, the inhomogeneities in the magnetic field (B) will give rise to variation in optical depths, and superposition of spectra with varying optical depths will broaden the resulting spectrum. Hence $\tau_{\rm SSA}^{'}$ is the effective optical depth coming from the superposition of the spectra of varying optical depths due to
different magnetic field in the model. $a^{'}$, $b^{'}$ denotes the temporal index of flux densities in the optically thick and thin phases respectively and $p^{'}$ is the power-law index of the electron energy distribution. 

We repeat the modeling to account for the effect of inhomogeneities in the FFA \citep{weiler2002} and SSA \citep{chandra2019,bjornsson2017} models. $K_{1}^{'}$, $K_{2}^{'}$, $\alpha^{'}$, $\beta^{'}$ and $\delta^{'}$ are the free parameters in the FFA model and $K_{1}^{'}$, $K_{2}^{'}$, $a^{'}$, $b^{'}$, $p^{'}$ and $\alpha^{''}$ are the free parameters in the SSA model.
The fits improve and the SSA model gives a better fit ($\chi_{\nu}^{2}$ = 0.9) than the FFA model ($\chi_{\nu}^{2}$ = 2.0). The best fit parameters for the inhomogeneous FFA model are $K_{1}^{'}=182.16\pm107.56$, $K_{2}^{'}=1919.19\pm1196.82$, $\alpha^{'}=-0.45\pm0.15$, $\beta^{'}=-1.95\pm0.19$, and $\delta^{'}=-3.90\pm0.21$ and for the inhomogeneous SSA model are $K_{1}^{'}=0.16\pm0.05$, $K_{2}^{'}=1366.26\pm989.02$, $a^{'}=1.57\pm0.18$, $b^{'}=2.07\pm0.19$, $p^{'}=2.33\pm0.33$, and $\alpha^{''}=1.40\pm0.22$. 
The 1.25, 0.75 and 0.40 GHz light curves of the best fit model peaks at $t_{\rm peak}$ = 138, 182 and 257 days with peak flux densities $F_{\rm peak}$ = 1.1, 0.9 and 0.6 mJy, respectively. The peak flux densities are derived by differentiating the best-fit solution to equation \ref{eqn:ssa-standard} and \ref{eqn:ssa-tau}.

\subsection{shock parameters}
The shock radius and magnetic field can be derived from the peak frequency ($\nu_{\rm peak}$) and peak flux density ($F_{\rm peak}$) of the SSA spectral energy distribution (SED) at a given time \citep{chevalier1998}.
If the emission structure is inhomogeneous, the peak flux density $F_{\rm peak}$ in a standard homogeneous SSA model needs to be replaced with $F_{\rm peak}/f_{\rm B,cov}$ to derive 
 various shock parameters \citep{bjornsson2017}. Thus the shock radius ($R$) and magnetic field ($B$) values in an inhomogeneous SSA model can be written as
\begin{eqnarray}
\label{eqn:Rp-general-inhomo}
R = R_{*} \times \left( f_{\rm B, cov} \right)^{\frac{-p-6}{2p+13}} 
\end{eqnarray}

\begin{eqnarray}
\label{eqn:Bp-general-inhomo}
B = B_{*} \times \left(f_{\rm B, cov} \right)^{\frac{2}{2p+13}} 
\end{eqnarray}

Where we have defined the shock radius and magnetic field in a standard homogeneous SSA model as $R_{*}$ and $B_{*}$, respectively. $R_{*}$ and $B_{*}$ can be expressed using the formulation presented in \citet{chevalier1998} as,

\begin{eqnarray}
\label{eqn:Rp-general}
R_{*} = \left[ \frac{6 c_{6}^{p+5} F_{\rm peak}^{p+6}D^{2p+12}}{(\frac{\epsilon_{\rm e}}{\epsilon_{\rm B}}) f (p-2) \pi^{p+5} c_{5}^{p+6} E_{\rm l}^{p-2}} \right]^{\frac{1}{(2p+13)}} \left( \frac{\nu_{\rm peak}}{2c_{1}} \right)^{-1} 
\end{eqnarray}

\begin{eqnarray}
\label{eqn:Bp-general}
B_{*} = \left[ \frac{36 \pi^{3} c_{5}}{(\frac{\epsilon_{\rm e}}{\epsilon_{\rm B}})^{2} f^{2}(p-2)^{2}c_{6}^{3} E_{\rm l}^{2(p-2)} F_{\rm peak} D^{2}}\right]^{\frac{2}{(2p+13)}} \left( \frac{\nu_{\rm peak}}{2c_{1}} \right)
\end{eqnarray}

In the above equations, $\nu_{\rm peak}$ is the peak frequency of the SED at a given time; 
  $\epsilon_{\rm e}$ and $\epsilon_{\rm B}$ denote the fraction of total energy density fed into relativistic electrons and magnetic fields respectively. $D$ denotes the distance to the source and $f$ is the volume filling factor of the radio emitting region taken as $f=0.5$ \citep{chevalier1998}. However the dependence of $R_{*}$ and $B_{*}$ on $f$ is weak. The value of $c_{1} = 6.265 \times 10^{18}$ in CGS units \citep{chevalier2017}. The constants $c_{5}$ and $c_{6}$ are tabulated for different values of $p$ in \cite{pacholczyk1970}. We use the values corresponding to $p=2.5$, the closest $p$ value in \cite{pacholczyk1970} with our best fit value ($p = 2.33$). $E_{\rm l}$ denotes the electron rest mass energy, i.e. 0.51 MeV.

If spatially resolved observations are available, the value of $f_{\rm B,cov}$ can be obtained  \citep{bjornsson2017}. Since the value of $f_{\rm B,cov}$ is not known for AT\,2018cow (although $f_{\rm B,cov}$ $<$ 1), 
we can only estimate $R_{*}$ and $B_{*}$ using equations  {\bf \ref{eqn:Rp-general-inhomo} and \ref{eqn:Bp-general-inhomo}}. Since $f_{\rm B,cov}$ $<$ 1, 
the actual shock radius will be larger than this estimate and magnetic field will be lower than this value. The mean shock velocity between $t=0$ and a particular age ($t$) can be estimated as $v = R/t$. However, we can only estimate $v_{*} = R_{*}/t$, which is a lower limit to the actual shock velocity.

\begin{deluxetable*}{cccccccccc}
\tablenum{2}
\tablecaption{Shock parameters of AT\,2018cow at multiple epochs.}
\tablewidth{0pt}
\tablehead{
\colhead{Parameters} & \multicolumn3c{$\epsilon_{\rm B}$=$\epsilon_{\rm e}$=0.33 } & \multicolumn3c{$\epsilon_{e}=0.1$, $\epsilon_{B}=0.01$} \\
\cline{2-7}
\colhead{} & \colhead{Day 138} & \colhead{Day 182} & \colhead{Day 257} & \colhead{Day 138} & \colhead{Day 182} & \colhead{Day 257}
}
\startdata
$R_{*}$ ($\times$10$^{16}$ cm)                & 6.12$\pm$0.71     &  9.28$\pm$1.07  &  14.36$\pm$1.67 & 5.38$\pm$0.62      &  8.16$\pm$0.95  & 12.64$\pm$1.47\\    
$v_{*}$ ($\times$ c)                            & 0.17$\pm$0.02     &  0.20$\pm$0.02  &  0.21$\pm$0.03  & 0.15$\pm$0.02      &  0.17$\pm$0.02  & 0.19$\pm$0.02 \\ 
$B_{*}$ ($\times$10$^{-1}$Gauss)               & 1.09$\pm$0.03     &  0.67$\pm$0.02  &  0.37$\pm$0.01  & 0.65$\pm$0.02      &  0.40$\pm$0.01  & 0.22$\pm$0.01 \\    
$E_{*}$ ($\times$10$^{49}$ erg)                & 0.08$\pm$0.03     &  0.11$\pm$0.04  &  0.13$\pm$0.04  & 0.67$\pm$0.24      &  0.88$\pm$0.31  & 1.01$\pm$0.36 \\    
$\dot{M}_{*}$ ($\times$10$^{-5}$M$_{\odot}$\,yr$^{-1}$)  & 0.41$\pm$0.13     &  0.27$\pm$0.09  &  0.17$\pm$0.05  & 4.82$\pm$1.60      &  3.15$\pm$1.05  & 1.96$\pm$0.65 \\ 
\enddata
\label{tab:shock-para}
\tablecomments { $R_{*}$,  $B_{*}$,  $E_{*}$ and $\dot{M}_{*}$ are the blast-wave radius, magnetic field, shock internal energy, and mass-loss rate defined in  eqns \ref{eqn:Rp-general}, \ref{eqn:Bp-general}, \ref{eqn:E-homo} and \ref{eqn:mass-loss-rate-homo}, and correspond to values of R, B, E and $\dot{M}$ for $f_{\rm B, cov}$=1, i.e.
a homogeneous SSA scenario. $v_{*}$ is the average shock velocity ($R_{*}/t$) between $t=0$ and the listed age under standard homogeneous assumption. }
\end{deluxetable*}

Following similar argument as above, the 
 shock internal energy ($E$) and the mass-loss rate  ($\dot{M}$) in an inhomogeneous scenario can be written as
\begin{eqnarray}
\label{eqn:E-inhomo}
E = E_{*} \times f_{\rm B, cov}^{\frac{-3p-14}{2p+13}}
\end{eqnarray} 

\begin{eqnarray}
\label{eqn:M-inhomo}
\dot{M} = \dot{M}_{*} \times f_{\rm B, cov}^{\frac{4}{2p+13}}
\end{eqnarray}  
 Here $E_{*}$ and $\dot{M}_{*}$ {\bf correspond} to the shock internal energy and mass-loss rate, respectively, in a standard homogeneous model. The value of $E_{*}$ can be obtained using the equation derived by \citet{soderberg2010a}. 
\begin{equation}
\label{eqn:E-homo}
    E_{*}=\frac{1}{\epsilon_{\rm B}}\left( \frac{B_{*}^{2}}{8\pi} \right) \frac{4}{3} \pi R_{*}^{3} 
\end{equation}
The $\dot{M}_{*}$ can be estimated from the magnetic field scaling relation \citep{chevalier1998}.
\begin{equation}
\label{eqn:mass-loss-rate-homo}
    \frac{B_{*}^{2}}{8\pi} = \epsilon_{\rm B} \frac{\dot{M_{*}}}{4 \pi R_{*}^{2} v_{\rm w}} v_{*}^{2}
\end{equation}

We derive the shock parameters assuming equipartition ($\epsilon_{\rm e}=$ $\epsilon_{\rm B}=0.33$) of energy between relativistic electrons and magnetic fields \citep{soderberg2010b}. While the shock radius and magnetic field are weakly dependent on the equipartition fraction, the shock internal energy and mass-loss rate changes significantly with $\epsilon_{\rm B}$. We also derive the shock parameters for non-equipartition values $\epsilon_{\rm B}=0.01$ and $\epsilon_{\rm e}=0.1$ for a comparison.  

\section{Results and Discussion}
\label{sec:results and discussion}
We determine $R_{*}$, $v_{*}$, $B_{*}$, $E_{*}$, and $\dot{M}_{*}$
corresponding to $f_{\rm B, cov} = 1$ using eqns \ref{eqn:Rp-general}, \ref{eqn:Bp-general}, \ref{eqn:E-homo} and \ref{eqn:mass-loss-rate-homo}. The actual $R$, $v$ and $E$ will be larger than these values and $B$, and $\dot{M}$ will be smaller.

\subsection{Properties of the shock}
The lower limit on the blast wave radius is $R_{*}$ $\sim$ 6.12 $\times$10$^{16}$ cm and 14.36 $\times$10$^{16}$ cm on $t$ = 138 and 257 days post explosion, respectively. The corresponding lower limit on mean shock velocity at these ages is $v_{*}$ $\sim$ 0.2c. 
The 3$\sigma$ upper limit on the shock radius on day 98 and 287 are $R$ $\sim$ $12.4 \times$ 10$^{16}$ cm and $R$ $\sim$ 58.4 $\times$ 10$^{16}$ cm respectively from VLBI observations \citep{bietenholz2019}. The corresponding upper limits on shock velocity are $v <$ 0.8c and 0.5c respectively. Thus the lower limits on shock radius and velocity derived from the uGMRT observations are consistent with the estimates from VLBI observations at similar epochs. 
 
 The mean shock velocity up to day 22 is reported as $v \sim$ 0.13c \citep{ho2019} and up to day 83 is $v \sim$ 0.1c \citep{margutti2019}. We estimate the mean shock velocity (using equation \ref{eqn:Rp-general}) to be $v$ $\sim$ 0.14c on day 102 from the peak flux density of the 5 GHz light curve \citep{mohan2019}. The lower limit on shock velocities derived from uGMRT observations are slightly large on $t =$ 257 days ($v_{*}$ $\sim$ 0.21 c). This marginal increase in shock velocity at later epochs could be indicative of a shock re-energization as it comes out of the dense circum-stellar shell which we discuss in \S \ref{sec:mass-loss}. 

The upper limits on the equipartition magnetic fields on day 138, 182, and 257 post-explosion are 0.11, 0.07, and 0.04 G respectively. These are similar to the magnetic fields seen in SNe\,Ibc \citep{chevalier2006}. 
The magnetic field strength in the shocked environment is expected to be $B \sim$ 10$^{4}$ G for models involving a neutron star \citep{lyutikov2019} which is significantly higher than the $B$ values derived from our analysis. Thus any scenarios involving a neutron star is less likely to be associated with AT\,2018cow.
 
The lower limit of the shock internal energy during $t =$ 138$-$257 days is $E_{*} \sim$(0.8$-$1.3) $\times$ 10$^{48}$ erg for $\epsilon_{\rm B}$ = $\epsilon_{\rm e}$ = 0.33.  The energy is sensitive to the choice of $\epsilon_{\rm B}$ value. The lower limit of the shock internal energy increases to $E_{*} \sim$ (0.7$-$1.0) $\times$ 10$^{49}$ erg for $\epsilon_{\rm B}$ = 0.01 and $\epsilon_{\rm e}$ = 0.1. The internal energies are comparable to that of most energetic SNe \citep{soderberg2010b,margutti2019}. 
 
 \subsection{Mass-loss rate}
 \label{sec:mass-loss}
We derive the mass-loss rate of the progenitor in a homogeneous SSA scenraio (using eqn \ref{eqn:mass-loss-rate-homo}) to be $\dot{M_{*}}$ $\sim$ (4.1$-$1.7)$\times$10$^{-6}$M$_{\odot}$\,yr$^{-1}$ for the shock parameters derived for $t =$ 138$-$257 days. We assume a wind velocity $v_{\rm w} =$ 1000 km\,s$^{-1}$ and $\epsilon_{\rm B} = 0.33$. The actual $\dot{M}$ values will be lower than these estimates.

\cite{ho2019} measure the $F_{\rm peak}$ and $\nu_{\rm peak}$ of the SSA spectrum on day 22 post-explosion and estimate the mass-loss rate to be $\dot{M}$ $\sim$ 4$\times$10$^{-4}$M$_{\odot}$\,yr$^{-1}$ \citep{ho2019}, two orders of magnitude greater than the $\dot{M}$ derived from uGMRT observations. The mass-loss rates probed by uGMRT observations are at an epoch of stellar evolution 19.3$-$45.7 years before the explosion, whereas the $\dot{M}$ derived at $t =$ 22 days correspond to $\sim$ 2.2 years prior explosion \citep{ho2019} for the assumed wind velocity. Thus the progenitor of AT\,2018cow goes through an enhanced phase of mass-loss close to the explosion.
 There are pieces of evidence in the literature for a dense shell of medium around AT\,2018cow and a possible cut-off in the density distribution from UVOIR \citep{perley2019} as well as radio observations \citep{ho2019}. \cite{ho2019} constrain the size of the dense CSM shell to be $R$ $\sim$ 1.7 $\times$ 10$^{16}$ cm from the substantial diminishing in the peak flux density of radio spectra. The uGMRT observations probe radii $\gtrsim$ 6 $\times$ 10$^{16}$ cm and are likely probing the material beyond the dense CSM region.
 
Assuming the surrounding medium to be composed of singly ionised hydrogen, we derive the electron number density $n_{\rm e}$ = $\frac{\dot{M}}{4 \pi r^{2}v_{\rm w} m_{\rm p}}$. The values of electron number densities are $n_{\rm e}$ $<$  33 cm$^{-3}$ at raduis $R_{*}$ $>$ 6.1 $\times$ 10$^{16}$ cm. At a radius $R \sim$ 7 $\times$ 10$^{16}$ cm, the $n_{\rm e}$ of SN\,2003bg is $\sim$ 43 cm$^{-3}$ \citep{soderberg2006-sn2003bg}.
 
\begin{figure}
\includegraphics[scale=0.36]{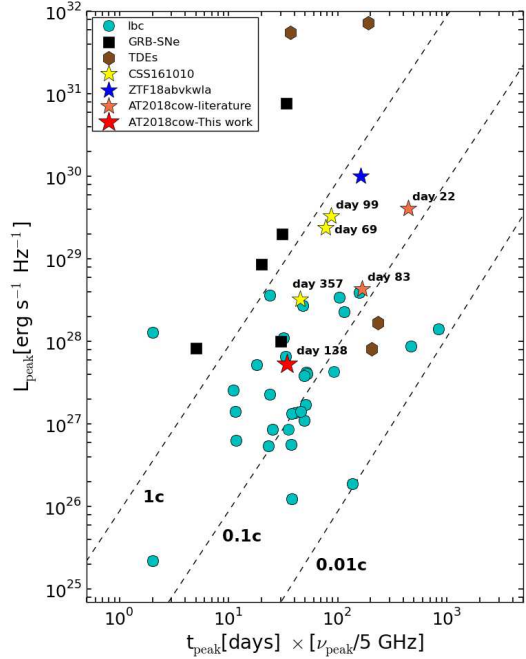}
\caption{The peak spectral luminosity and peak time of AT\,2018cow from uGMRT observations are plotted along with those of other energetic transients. Other FBOTs with radio detections, CSS161010 and ZTF18abvkwla, are also marked for a comparison. The dotted line denotes the mean shock velocity in a SSA scenario. References:  \citep{kulkarni1998,vanderhorst2008,soderberg2010b,margutti2013-sn2010bh,chevalier2017,chandra2019,nayana2020,alexander2016,alexander2017,cenko2012,berger2012,zauderer2011,ho2019,margutti2019,ho2020,deanne2020} and references therein.}
\label{fig:comparison}
\end{figure}
\section{A comparison between AT\,2018cow with other energetic transients}
\label{sec:comparison}
We compare the radio properties of AT\,2018cow with those of other energetic transients in Fig \ref{fig:comparison}. The mean shock velocities in an SSA scenario are plotted as dotted lines \citep{chevalier1998}. While AT\,2018cow shows unusually high radio luminosity and peak frequency at early time \citep[day 22;][]{ho2019}, the peak spectral luminosity of AT\,2018cow from the uGMRT observation is comparable to that of other type Ib/c SNe (SNe\,Ibc) at similar epochs. This indicates that the high $L_{\rm peak}$ and $\nu_{\rm peak}$ of AT\,2018cow is  detected due to the early observation campaign at sub-mm frequencies. Other SNe\,Ibc could as well show high $L_{\rm peak}$ and $\nu_{\rm peak}$ had those been observed immediately after the explosion. 
The peak spectral luminosity of AT\,2018cow at $t_{\rm peak}$ = 22 \citep{ho2019} and 83 days \citep{margutti2019} are also marked in Fig \ref{fig:comparison} for a comparison. The temporal evolution of AT\,2018cow in the $L_{\rm peak}$-$t_{\rm peak}$ diagram is roughly through a constant velocity line. The shock velocity of AT\,2018cow is relatively low compared to those of other FBOTs, $v$ $\geq$ 0.5c for CSS161010 \citep{deanne2020} and  $v$ $\geq$ 0.3c for ZTF18abvkwla \citep{ho2020}. 
\cite{deanne2020} reports evident deceleration in the shock velocity of CSS161010 from $v$ = 0.55 $\pm$ 0.02 c to $v$ = 0.36 $\pm$ 0.04 c during day 99 to 357 post-explosion. Such a deceleration is not seen in AT\,2018cow up to day 257 post-explosion. The three FBOTs with multi-wavelength follow-up observations show diverse properties. A bigger sample of these events will reveal the variety in their intrinsic properties.

\section{Summary}
\label{sec:summary}
We present uGMRT observations of AT\,2018cow at 1.25, 0.75, and 0.40 GHz during 11$-$570 days post the event. The peak luminosity at 1.4 GHz is 5.3 $\times$ 10$^{27}$ erg\,s$^{-1}$\,Hz$^{-1}$. While AT\,2018cow shows remarkable luminosity in the sub-mm bands at early times \citep[$t \sim$ 22 days;][]{ho2019}, the luminosity of the transient at late times is very similar to that of energetic SNe\,Ibc. The uGMRT observations are best represented by a self-absorbed inhomogeneous synchrotron emission model. Assuming the source covering factor to be unity, we estimate the shock radius, magnetic field, mass-loss rate of the progenitor, and shock internal energy. The actual shock radius will be larger and the magnetic field will be smaller than these values since the radio-emitting region is inhomogeneous. The mass-loss rate and energy estimates will also be upper and lower limits respectively. We derive the lower limit of shock radius to be $R_{*}$ $\sim$ (6.12$-$14.36)$\times$10$^{16}$ cm during $t =$ 138$-$257 days, consistent with VLBI observations covering similar epochs \citep{bietenholz2019,mohan2019}. The lower limit on the average shock velocity on $t =$ 257 day is $v_{*}$ $\sim$ 0.21 c for $\epsilon_{\rm B} = \epsilon_{\rm e} = 0.33$ and $v_{*}$ $\sim$ 0.19 c for $\epsilon_{\rm B} = 0.01,\,\epsilon_{\rm e} = 0.1$, indicating that the fast-moving ejecta from the event do not experience any deceleration up to $t =$ 257 days. The upper limit on the equipartition magnetic field on $t =$ 138$-$257 days is in the range (0.11$-$0.04) G, much smaller than the expected magnetic field in models involving a neutron star \citep[10$^{4}$ G;][]{lyutikov2019}. The upper limit on the mass-loss rate of the progenitor is $\dot{M_{*}}$ $\sim$ (4.1$-$1.7) $\times$ 10$^{-6}$ M$_{\odot}$\,yr$^{-1}$ for the limits on the shock parameters derived for $t =$ 138$-$257 days, $\sim$ 10$^{2}$ times lower than the mass-loss rates derived from early ($t =$22 days) sub-mm observations \citep{ho2019}. This is consistent with the speculation of a dense circumstellar shell in the vicinity of AT\,2018cow \citep{ho2019,perley2019} if the uGMRT observations are probing the material beyond this dense shell. Our results reveal the importance of low-frequency radio observations to probe the  environments of FBOTs at later epochs.
\acknowledgments
We thank the referee for the critical comments, which helped to improve the manuscript significantly. We thank Rupak Roy and Varun Bhalerao for their support at various stages of this work. P.C. acknowledges support from the Department of Science and Technology via the SwaranaJayanti Fellowship award (file no.DST/SJF/PSA-01/2014-15). We acknowledge the support of the Department of Atomic Energy, Government of India, under project no. 12-R\&D-TFR-5.02-0700. We thank the staff of the GMRT that made these observations possible. The GMRT is run by the National Centre for Radio Astrophysics of the Tata Institute of Fundamental Research.

%

\facilities{uGMRT}


\software{CASA \citep{mcmullin2007},  
          Python-scipy
          }







\end{document}